\documentstyle[12pt,twoside,fleqn,espcrc1]{article}

\newcommand{\AmS}{{\protect\the\textfont2A\kern-.1667em
\lower.5ex\hbox{M}\kern-.125emS}}
\renewcommand{\textfraction} {.01} 
\renewcommand{\topfraction} {.99} 
\newcommand {\be} {\begin{eqnarray}} 
\newcommand {\ee} {\end{eqnarray}} 
\input psfig.sty
\begin{document} 
 \topmargin -0.3in
\oddsidemargin -0.50cm 
\evensidemargin 0cm 
\textwidth 6.5in
\textheight 8.5in 
\parindent 1.2cm 
\renewcommand{\textfraction} {.01} 
\renewcommand{\topfraction} {.99}

\pagestyle{empty}
\Huge{\noindent{Istituto\\Nazionale\\Fisica\\Nucleare}}

\vspace{-3.9cm}

\Large{\rightline{Sezione di ROMA}}
\normalsize{}
\rightline{Piazzale Aldo  Moro, 2}
\rightline{I-00185 Roma, Italy}

\vspace{0.65cm}

\rightline{INFN-1263/99}
\rightline{September 1999}

\vspace{1.cm}

\begin{center}{\large\bf A light-front
description of electromagnetic form factors for $J \leq \frac{3}{2}$ hadrons}
\end{center}
\vskip 1em
\begin{center}  E. Pace$^a$, G. Salm\`e$^b$, F. Cardarelli and S.
Simula$^c$\end{center}

\noindent{$^a$\it Dipartimento di Fisica, Universit\`a di Roma
"Tor Vergata", and Istituto Nazionale di Fisica Nucleare, Sezione
Tor Vergata, Via della Ricerca Scientifica 1, I-00133, Rome,
Italy} 

\noindent{$^b$\it Istituto Nazionale di Fisica Nucleare, Sezione
di  Roma I, P.le A. Moro 2, I-00185 Rome, Italy} 

\noindent{$^c$\it Istituto Nazionale
di Fisica Nucleare, Sezione Roma III, Via della Vasca Navale 84, I-00146 Roma,
Italy}

\vspace{2.cm}

\begin{abstract}A review of the hadron electromagnetic form factors obtained 
in a light-front constituent quark model, based on the eigenfunctions of a mass
operator, is presented. The relevance of different components in the q-q 
interaction for the description of hadron experimental form factors is analysed. 
\end{abstract}

\vspace{7.5cm}
\hrule width5cm
\vspace{.2cm}
\noindent{\normalsize{Proceedings of "Nucleon 99",  Frascati, June 1999. 
To appear in {\bf Nucl. Phys. A}}} 

\newpage


\pagestyle{plain}

\section{INTRODUCTION}

Within a relativistic light-front (LF) constituent-quark ($CQ$) model,
 we performed an extended investigation of elastic and transition
electromagnetic (e.m.) hadron form factors~\cite{CAR_N} in the 
momentum transfer region relevant for the experimental research programme at
TJNAF \cite{TJNAF}. The main features of the model are: i) 
eigenstates of a mass operator which reproduces a large part of the hadron
spectrum; ii) a one-body current operator with phenomenological Dirac 
and Pauli form factors for the $CQ$'s.
 The $CQ$'s are assumed to interact via the $q - q$ potential of
Capstick and Isgur ($CI$) \cite{CI86}, which includes a linear confining term 
and an effective one-gluon-exchange ($OGE$) term. 
The latter produces a huge amount of high-momentum components in the baryon wave
functions \cite{CAR_N} and contains a central Coulomb-like potential, a
spin-dependent part, responsible for the hyperfine splitting of baryon masses,
and a tensor part. A comparable amount of high momentum components was obtained
\cite{CAR_N}(f) with the $q - q$ interaction based on the exchange of the
pseudoscalar Goldstone-bosons \cite{Gloz}. This fact suggests that the hadron
spectrum itself dictates the high momentum behaviour in hadron wave functions.
In this paper a review of our results for the elastic and
transition form factors for $J \leq \frac{3}{2}$ hadrons is presented
\cite{CAR_N}(a-g).

\section{ELECTROMAGNETIC HADRON FORM FACTORS IN THE LF DYNAMICS}

\indent In the $LF$ formalism the space-like e.m. form factors can be
related to the matrix elements of the {\em plus} component of the e.m. current,
${\cal{I}}^+ = {\cal{I}}^0 + {\cal{I}}_z$, in the reference
frame where $q^+ = q^0 + {q}_z = P^{+}_f - P^+_i = 0$.
We have evaluated elastic and transition form factors (f.f.) 
by assuming the ${\cal{I}}^+$ component of the e.m. current to be the sum
of one-body $CQ$ currents  \cite{CAR_N}, i.e.
    ${\cal{I}}^+(0) \approx \sum_{j=1}^3 ~ I^+_{j}(0) = \sum_{j=1}^3 ~ \left (
    e_j \gamma^+ f_1^j(Q^2) ~ + ~ i \kappa_j {\sigma^{+ \rho} q_{\rho}
    \over 2 m_j}f_2^j(Q^2) \right )$
\noindent with $e_j$ ($\kappa_j$) the charge (anomalous magnetic moment) of
the j-th quark, and $f_{1(2)}^j$ its  Dirac (Pauli) form factor.
We studied first pion and nucleon elastic form factors
and showed that an effective one-body e.m. current, with a suitable choice for
the $CQ$ form factors, is
able to give a coherent description of pion and nucleon experimental data
\cite{CAR_N}(a).

\begin{figure}[t]
\psfig{figure=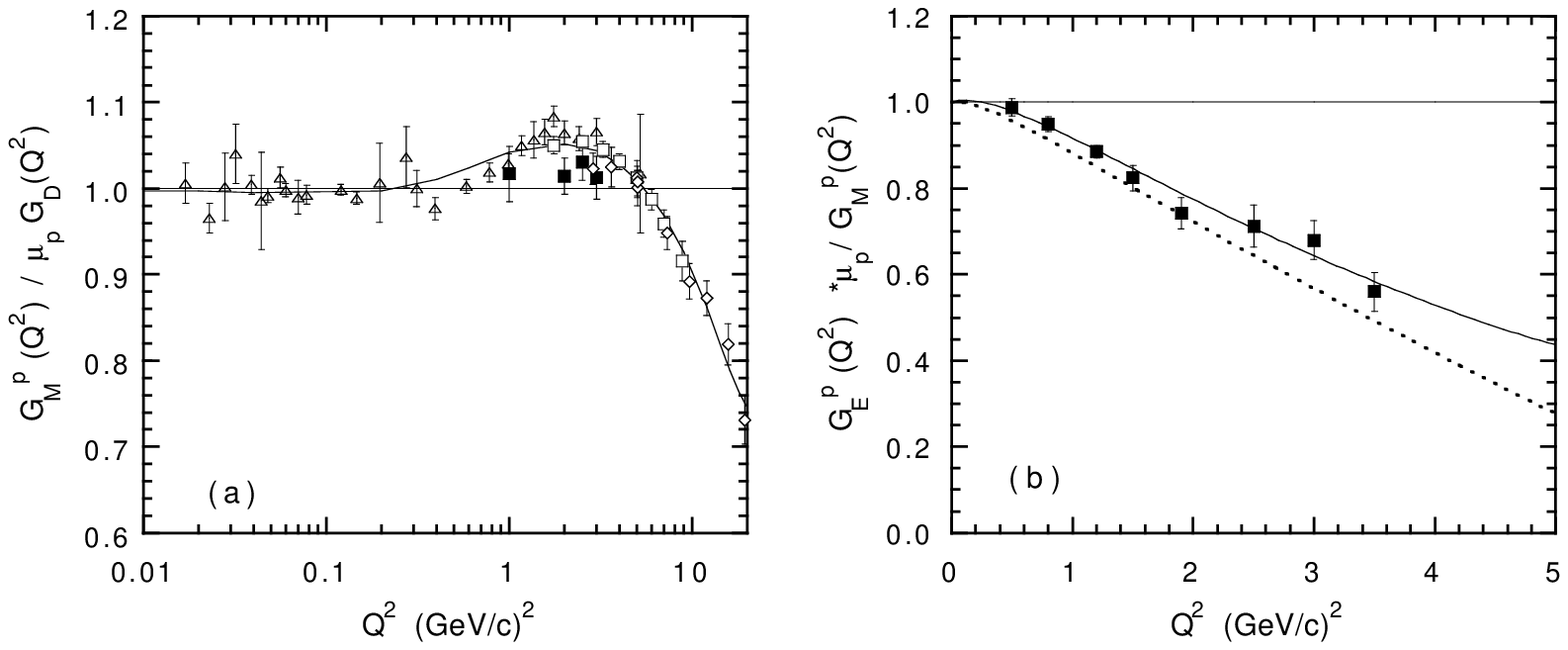,bbllx=11mm,bblly=215mm,bburx=0mm,bbury=285mm}
Figure 1.  { The  proton magnetic form factor 
$G_M^p(Q^2)/\mu_pG_D(Q^2)$ (a), and the ratio $G_E^p(Q^2)\mu_p/G_M^p(Q^2)$ (b) 
vs. $Q^2$ ($G_D(Q^2) = (1+Q^2/0.71)^{-2}$. In (b) the solid and dashed lines represent
our results with and without $CQ$ form factors, respectively. Experimental
data in (a) are as quoted in Ref.~\cite{CAR_N}(a); experimental
data in (b) are from Ref. \cite{Perdri}. } 
\end{figure} 

\indent In this paper our fit of the $CQ$ form factors is updated to describe
the most recent data for the nucleon f.f., in particular for the ratio
$G_{Ep}\mu_p/G_{Mp}$ \cite{Perdri}.

In Fig. 1 the elastic proton form factors are shown, in order to illustrate 
the
high quality fit one can reach (a fit of the same quality is obtained for the
neutron and the pion as well). It is interesting to note that, while effective
CQ f.f. are required to describe the nucleon f.f., the experimental data
for the ratio $G_E^p \mu_p/ G_M^p$ can also be reproduced by the current with
pointlike CQ's (see dashed line in Fig. 1 (b)). 
Therefore this ratio appears to be directly linked to the structure of the
nucleon wave function.  

\section{NUCLEON-RESONANCE TRANSITION FORM FACTORS}

Once the $CQ$ form factors have been determined, we can obtain 
{\em {parameter-free}} predictions for the nucleon-resonance transition form
factors.
\begin{figure}
\psfig{figure=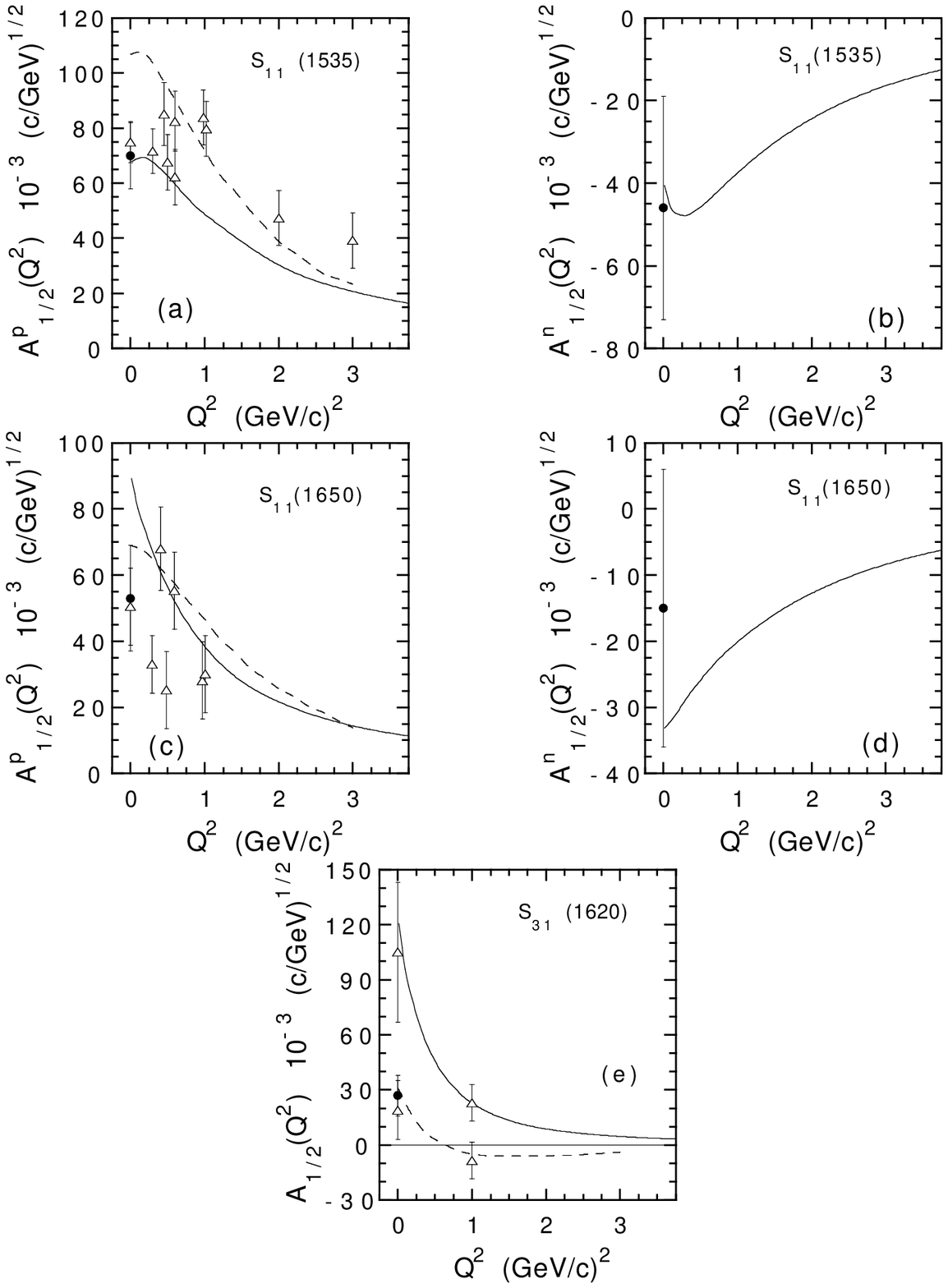,bbllx=5mm,bblly=93mm,bburx=0mm,bbury=285mm}
Figure 2.  { The transverse helicities $A_{1/2}^{p(n)}$ for the nucleon 
transitions
$p \rightarrow S_{11}(1535)$ (a); $n \rightarrow S_{11}(1535)$ (b);
$p \rightarrow S_{11}(1650)$ (c); $n \rightarrow S_{11}(1650)$ (d);
$p \rightarrow S_{31}(1620)$ (e), vs. $Q^2$.
The solid and dashed lines represent our calculations and the results of a
non-relativistic $CQ$ model \cite{Aiello}, respectively. Solid dot: PDG '96
\cite{DEXP}; triangles: data analysis from \cite{Burkert}. }   
\end{figure}
\begin{figure}
\psfig{figure=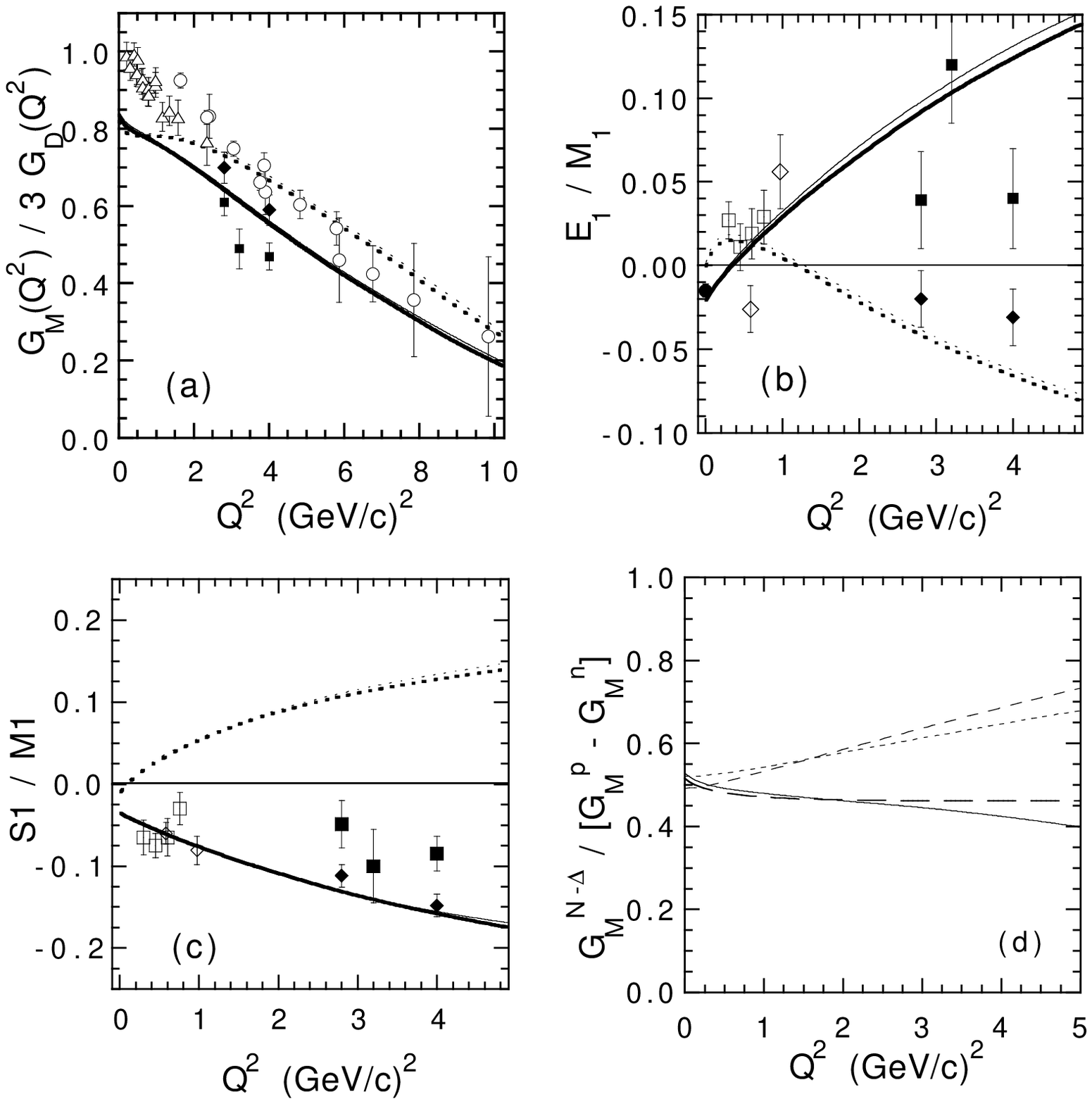,bbllx=11mm,bblly=120mm,bburx=0mm,bbury=285mm}
Figure 3.  {(a) The $N-\Delta$ transition magnetic form factor $G_M^{N-\Delta}
\left (Q^2 \right ) / 3G_D \left (Q^2 \right )$, vs $Q^2$. Solid and dotted 
lines are the results of prescriptions i) and ii) of Ref. \cite{CAR_N}(d).
Thick and thin lines correspond to the full calculation with the
$CI$ interaction \cite{CI86} and to the contribution of the $S$-wave 
in the $\Delta$ eigenstate, respectively. 
Triangles: Ref. \cite{D_EXP}(a); open dots: Ref. \cite{D_EXP}(b); full
squares: analysis of Ref. \cite{D_EXP}(c); full diamonds: Ref. \cite{D_EXP}(d). 
-  (b) The same as in (a), but for $E_1/M_1$. Full dot: PDG \cite{DEXP};
open diamonds: Ref. \cite{EXP}(a); open squares: Ref. \cite{EXP}(b); 
full squares: analysis of Ref. \cite{D_EXP}(c); full diamonds:
Ref. \cite{D_EXP}(d). - (c) The same as in (b), but for $S_1/M_1$. - (d)
The ratio $G_M^{N-\Delta}\left (Q^2 \right )/(G^p_M \left (Q^2 \right )-G^n_M
\left (Q^2 \right ))$ vs $Q^2$.  Solid line:  our calculation (prescription 
ii) of Ref. \cite{CAR_N}(d)) with the
$CI$ baryon eigenfunctions and $CQ$ form factors; dashed line: the same as
the solid line, but without $CQ$ form factors; short-dashed line: the
same as the dashed line, but with the baryon eigenfunctions
corresponding to the spin-independent part of the $CI$ interaction
\cite{CI86}; dotted line: the same as the dashed line, but retaining
only the confining part of the $CI$ potential.}   
\end{figure}

In Fig. 2 our evaluations of the helicity amplitude
$A_{1/2}$ are shown for $N \rightarrow S_{11}(1535)$, $S_{11}(1650)$ and
$S_{31}(1620)$, and compared with the results of a non-relativistic model
\cite{Aiello}. In the case of $S_{31}(1620)$ the results for $p$ and $n$
coincide (as for $P_{33}(1232)$), since only the isovector part of the CQ
current is effective. Our predictions yield an overall agreement with available
experimental data for the $P$-wave resonances and show a sizeable sensitivity to
relativistic effects, but more accurate data are needed to reliably discriminate
between different models.

Our parameter-free predictions for the ${N-\Delta(1232)}$ 
transition form factors, obtained using the prescriptions i) and ii)
defined in
\cite{CAR_N}(d), are compared with existing data in Fig. 3 (a),(b),(c). 
In Fig. 3 (d) the ratio between $G_M^{N-\Delta}\left (Q^2 \right )$ and
the isovector part of the nucleon magnetic form factor, $G^p_M \left
(Q^2 \right )-G^n_M \left (Q^2 \right )$, is shown to be largely
insensitive to the presence of $CQ$ form factors, whereas it is
sharply affected by the spin-dependent part of the $CI$ potential,
which is generated by the chromomagnetic interaction. It can
clearly be seen that: i)  the effect of the tiny $D$-wave component
($P_D^{\Delta} = 1.1 \%$), and then of the tensor part in the $q-q$ interaction,
is small for $G_M^{N-\Delta}$, as well as for $E_1/ M_1$ and $S_1/M_1$ 
 (the $L=0$ component gives non-zero values of $E_1 / M_1$
and $S_1/M_1$, because of the relativistic nature of our calculation); ii) 
 the effect of the spin-dependent part of the $q-q$
interaction, which is responsible for the ${N-\Delta}$ mass splitting and for
the different high momentum tails of the $N$ and $\Delta$ wave functions, is
essential to reproduce the faster-than-dipole fall-off of $G_M^{N-\Delta}
\left (Q^2 \right )$ (see also Ref. \cite{CAR_N}(f)).
Both these results do not depend on the prescriptions used to extract the
${N-\Delta}$ transition form factors.

\end{document}